
\vskip 24pt
\magnification=\magstep1
\hsize=6truein
\rightline{CU-TP-640}

\vskip 24pt

\leftskip=.25in
\centerline{\tfont \bf UNITARITY AND THE BFKL POMERON\footnote{*}{\sevenrm
This work
is supported in part by the Department of Energy Under Grant
DE-FG02-94ER40819}}
\vskip 10pt
\centerline{A.H. Mueller}
\centerline{\it Department of Physics, Columbia University, New York, N.Y.
10027}
\vskip 24pt
{\narrower\smallskip High Energy onium-onium scattering is calculated
as a function of impact parameter in the one and two pomeron exchange
approximation.  Difficulties with using the multiple scattering series to
unitarize single pomeron exchange at high energy are noted.  An operator
formalsim which sums all numbers of pomeron exchange is given.  A toy model
which has a similar operator structure at high energy as QCD is presented and
the S-matrix is evaluated.  Estimates of the energies and impact parameters at
which blackness occurs in onium-onium scattering are given.  It is emphased
that the problem of unitarity in high energy onium-onium scattering can be
solved in a purely perturbative context, with a non-running coupling if the
onium is heavy enough.\smallskip}
\vskip 15pt
\baselineskip=18pt
\centerline{\tfont\bf 1.  Introduction}
\vskip10pt

The Balitsky, Fadin,Kuraev and Lipatov (BFKL)[1-3]  pomeron applies to
processes which are at the same time hard processes and high energy processes.
On the conceptual level perhaps the simplest process where the BFKL pomeron
applies is in high energy onium-onium scattering.  For a sufficiently heavy
onium state perturbation theory naturally applies while for high enough energy
the essential, and interesting, aspects of having to deal with collisions
of projectiles consisting of a large numbr of partons are present.

In earlier papers[4-5] a picture has been developed for high energy
scattering in which a high energy onium state looks like a collection of
independent color dipoles of various sizes. (For a somewhat related discussion
see ref.6
and 7.) In a high energy onium-onium elastic scattering the scattering occurs
due to a
dipole
in one onium state scattering off a dipole in the other onium state by means of
the
exchange
of two gluons.  The BFKL pomeron amplitude was recovered in this dipole picture
with the onium-onium cross section due to a single dipole scattering given by

$$\sigma^{(1)}(Y) = 16\pi^2R^2\alpha^2{e^{(\alpha_P-1)Y}\over [{7\over 2}\alpha
N_c\zeta(3)Y]^{1/2}}\eqno(1)$$

\noindent where  R  is the onium radius and $Y=\ell n\  s/M^2$ with\ M\ the
onium
mass and\ s\ the square of the center of mass energy of the collision.  The
cross section due to the independent interaction of two dipoles in one onium
with two dipoles in the other onium was also found to be

$$\sigma^{(2)}(Y) = c {R^2[\sigma^{(1)}(Y)]^2\over [{7\over 2}\alpha
N_c\zeta(3)Y]^2}$$

\noindent with\ c\ a constant.

It is perhaps surprising that $\sigma^{(2)}$ is not just a constant times
$R^2[\sigma^{(1)}]^2$ since the BFKL pomeron is a fixed cut in the angular
momentum plane.  However, the explanation is apparent in impact parameter space
where one finds, for $b/R >> 1,$

$${d\sigma^{(1)}\over d^2b} =  8\pi\alpha^2 {R^2\over b^2} \ell
n(b^2/R^2)\ {e^{(\alpha_P-1)Y-{a\over 2}\ell n^2(b^2/R^2)}\over [{7\over
2}\alpha N_c\zeta(3)Y]^{3/2}}$$

\noindent from (10) with a given by (9).  Thus $\sigma^{(1)}$ is determined
mostly by impact parameters much larger than  R.  We shall see that the
dominant contribution to $\sigma^{(1)}$ comes from the interaction of dipoles
dilutely spread over a large radius in impact parameter space.  Because the
dipoles are dilute they do not contribute very much to $\sigma^{(2)}$ so that
$\sigma^{(2)}$ is determined from the interactions of dipoles no further than
R  from the center of the heavy quark-antiquark pair.

At sufficiently large  Y  one does not expect (1) to be a good representation
of the total onium-onium cross section.  Indeed, the growth in Y of
$\sigma^{(2)},$ and of the ``cross sections'' corresponding to higher numbers
of dipole interactions, is much stronger than that of $\sigma^{(1)}.$  However,
we have just seen above that unitarity corrections are likely to be much
simpler when viewed in impact parameter space.  To that end single pomeron
exchange is examined, in terms of dipole scattering, in impact parameter space,
in sect.2.  In the appendix double pomeron exchange is evaluated in impact
parameter space and a curious phenomena is found.  For $b/R >> 1$ one finds
${d\sigma^{(2)}\over d^2b} >> [{d\sigma^{(1)}\over d^2b}]^2.$  The cause of
this
is that ${d\sigma^{(2)}\over d^2b}$ is dominated by fluctuations in the onium
wavefunction which have little probability but which have a very large
two-dipole cross section.  This unsettling result means that the dominant term
in the two pomeron exchange is coming from  different parts of the onium
wavefunction than determine single pomeron exchange.  Thus, the dominant part
of
two pomeron exchange is not serving to unitarize single pomeron  exchange.

In sect.3, an operator formalism for onium-onium scattering is
developed and an expression for the S-matrix is given in eq.(32).  The S-matrix
given by (32) in principal has all numbers of pomeron exchange
included.  As an operator expression it looks like an eikonal
approximation, however, when evaluated it does not correspond to
an eikonal approximation as the explicit calculations in sect.2
and the appendix show.  Eq.(32) gives an S-matrix obtained by
evaluating  the wavefunction in the large $N_c$ limit and  in the
leading logarithmic approximation in longitudinal momentum. The
S-matrix (32) is given by the sum of all numbers of dipole-dipole
scatterings in the onium-onium collision.  Since each
dipole-dipole scattering is down by $1/N_c^2$ the S-matrix has
all terms of order $1/N_c^2$ included.  Arguments are given to
the end that higher order $1/N_c^2$ corrections to the
wavefunction need not be included to get a good approximation to
the S-matrix.

In sect.4, a toy model is given which has a structure for scattering similar to
that of QCD but which has no transverse dimensions.  This allows an exact
evaluation of the wavefunction in which the number of partons is distributed
in a KNO[8] manner.  The formula for the S-matrix given by (47) is structurally
identical to (32), but without the complexity of the many degrees of freedom
due
to  transverse dimensions.  The multiple scattering series is divergent but
Borel summable.  The multiple scattering series given by (32) for
four-dimensional QCD is also likely to be a divergent series with N!'s
appearing in the scattering of  N  dipoles.

In sect.5 we give qualitative estimates of the S-matrix in high energy
onium-onium scattering including unitarity corrections.  We are able to follow
the behavior of the S-matrix, and the resulting onium-onium cross section, to
energies somewhat beyond that where the onium-onium cross section is black
out to impact parameters as large as the diffusion radius of the bFKL pomeron.
As energies increase and blackness goes beyond the diffusion radius the first
strong modifications of the BFKL formula for the total cross section are
evident.  At lower energies there are strong unitarity corrections at small
impact parameters, but the BFKL formula for the total cross section is
esssentially unmodified.

It should be emphasized that the S-matrix given by (32) is determined purely
from perturbation theory.  That is, unitarity is realized by a perturbative,
even a fixed coupling, calculation.  The truly nonperturbative phenomena which
are involved in trying to understand the high field strength saturation[9]
limit
only occur at higher energies after blackness of  the cross section has set in.

Although we have not been able to evaluate the S-matrix given by (32)
analytically, it should be emphasized that a numerical evaluation of (32) is in
principle straightforward. Such an evaluation would necessitate constructing
the square of the many dipole wavefunction by a Monte Carlo evaluation of the
branching process giving that wavefunction.  Then one would evaluate $e^{-f}$
for each Monte Carlo generated event, and finally average over the Monte Carlo
events.

Finally, we note that Li and Tan[10] have recently solved the high energy
behavior
of scattering in 2+1-dimensional QCD.  In their calculation they find that
single pomeron exchange leads to a logarithmically decreasing cross section at
high energy.  Muultiple scattering terms are asymptotically smaller.  Li and
Tan
also introduce an impact parameter description of the scattering which is very
similar to the one used in the present paper.
\vfill
\eject
\centerline{\bf 2. Single Pomeron Exchange}
\vskip 10pt

In this section single pomeron exchange in onium-onium scattering will be
calculated.  The calculation follows that given in ref.5,  but here the result
is given for onium-onium scattering at a definite impact parameter.  A(Y,b) can
be written as

$$A =\ -\ i \int d^2 x_{01}d^2x_{01}^\prime\int_0^1 dz dz^\prime
\Phi(x_{01},z)\
\Phi(x_{01}^\prime, z^\prime)F\eqno(2)$$

\noindent where for a single pomeron exchange $F=F^{(1)}$ with

$$F^{(1)}= -4\pi\alpha^2 \int_0^\infty {d\ell\over \ell^3}\ {dx\over
x} {dx^\prime\over x^\prime}
 \int d^2b_1(1-J_0(\ell x))(1-J_0(\ell
x^\prime))$$
$$\cdot n(x_{01},x,Y/2,b_1)n(x_{01}^\prime,x^\prime,Y/2, \vert\b b-\b
b_1\vert).\eqno(3)$$

$\Phi(x_{01},z)$ is the square of the heavy quark-antiquark part of the onium
wavefunctions with $x_{01}$ the transverse coordinate separation between the
heavy quark and antiquark, and $z_1$ is the longitudinal momentum fraction of
the antiquark.  $n(x_{01},x,Y/2,b_1)$ is the number density of dipoles of size
x  at a transverse distance $b_1$ from the center of the heavy quark-antiquark
pair where the momentum fraction of the softest of the two gluons involved in
making up the dipole is greater than $e^{-Y/2}.$ (Recall  that in the large
$N_c$ limit one can view each gluon in the onium wavefunction as a
quark-antiquark pair.  The dipole  x  is made of a quark part of one gluon and
an antiquark part of another gluon.)  In the center of the mass system, eq.(3)
expresses onium-onium scattering as the interaction, due to two gluon exchange,
of a right-moving dipole in one onium state with a left-moving dipole in the
other onium state.  Eqs.(2) and (3) are the direct analogues of eqs.(8) and
(20) of ref.5, but where now we have introduced scattering at a definite impact
parameter.  We anticipate $x,x^\prime << b_1,\vert \b b-\b b_1\vert$ so that
the centers of the dipoles $\b x$ and $\b x^\prime$ can be taken at the same
point.

The relationship between the impact parameter number densities and those in
momentum space is

$$n(x_{01},x,Y,b) = \int {d^2\b q\over (2\pi)^2} e^{-i\b q\cdot \b
b}n(x_{01},x,Y,q)\eqno(4)$$

where

$$n(x_{01},x,Y,q)=\int {d\nu\over
2\pi}E_q^{0\nu^*}(x_{01})E_q^{0\nu}(x)({x_{01}\over
x})exp[{2\alpha N_c\over \pi}\chi(\nu)Y]\eqno(5)$$

with

$$\chi(\nu)=\psi(1)-{1\over 2}\psi({1\over 2}-i\nu)-{1\over 2}\psi({1\over
2}+i\nu)\eqno(6)$$

\noindent and

$$E_q^{0\nu}(x)={2i\nu\over \pi}x^{2i\nu}\int {d^2\b R\  e^{i\b q\cdot \b
R}\over
[\vert \b R-\b x/2\vert \vert\b R + \b x/2\vert]^{1+2i\nu}}\eqno(7)$$

\noindent for $\nu <<1,$ the region of $\nu$ which dominates the
integral in (5) when $\alpha N_cY >>1.$

When $b/x, b/x_{01} >> 1$ the integral in (4) can be done, using (5)-(7),
and yields

$$n(x_{01},x,Y,b) = {x_{01}\over 4 x b^2}\ell n({b^2\over
x_{01}x}){exp\{(\alpha_P-1)Y-{a(Y)\over 2}\ell n^2({b^2\over x_{01}x})\}\over
[{7\over 2}\alpha N_c\zeta(3)Y]^{3/2}}\eqno(8)$$

\noindent with $\alpha_P-1={4\alpha N_c\over \pi}\ell n 2$ and where

$${\bigg a}(Y)=[7\alpha N_c\zeta(3)Y/\pi]^{-1}.\eqno(9)$$

\noindent Substituting (8) into (3) one finds

$$F^{(1)}(Y,b) = - {\pi\alpha^2x_{01}x_{01}^\prime\ell
n({b^2\over x_{01}x_{01}^\prime})\over [{7\over 2}\alpha
N_c\zeta(3)Y]^{3/2}b^2}
exp\{(\alpha_P-1)Y-{a\over 2}\ell n^2({b^2\over
x_{01}x_{01^\prime}})\}.\eqno(10)$$

\noindent We note that the forward scattering amplitude,  $F^{(1)}(Y),$ is
given by

$$F^{(1)}(Y)=\int d^2bF^{(1)}(Y,b)= - {2\pi x_{01}x_{01}^\prime\alpha^2
e^{(\alpha_P-1)Y}\over \sqrt{{7\over 2}\alpha N_c\zeta(3)Y}}\eqno(11)$$

\noindent in agreement with eq.(26) of ref.5.

Comparing (10) and (11) it is clear that the dominant part of the total cross
section, in
the one pomeron exchange approximation, comes from impact parameters  b  given
by

$$\ell n({b^2\over x_{01}x_{01}^\prime}) \sim {\sqrt{14\alpha
N_c\zeta(3)Y/\pi}}.$$

\noindent Thus the cross section is dominated by impact parameters much larger
than the
average onium radius, R.  From (8) one sees that the dipole number density
become small
when $b^2$ becomes large.  Even when b/R is not too large it is easy to check
from (2)
that
scattering takes place through the interaction of dipoles of size $x \approx
x^\prime << b$
and that these small dipoles are spread rather dilutely over the area of
scattering so long
as $\alpha Y$ is not too large.
\vskip 10pt

\centerline{\bf 3.  Operator Formulation of High Energy Onium-Onium Scattering}
\vskip 6pt

In this section, we give a formulation of onium-onium scattering in terms of
operators $a(
\b b,\b x)$ and  $a^+(\b b,\b x)$ which destroy and create a dipole of size $\b
x$ and whose center is at $\b b.$  As in ref.4 let  Z  be the generating
functional for
dipoles starting with a dipole of separation $\b x_{01} = \b x_1-\b x_0$ whose
center is
at
$\b b_0={1\over 2} (\b x_0+\b x_1).$  Then

$$Z(\b b_0,\b x_{01},Y,u) = u(\b b_0,\b x_{01})exp\{-{2\alpha N_c\over \pi}\ell
n({x_{01}\over \rho})Y\}
+{\alpha N_c\over \pi^2}\int_\rho {x_{01}^2 d^2x_2\over x_{20}^2 x_{12}^2}$$
$$\cdot exp\{-{2\alpha N_c\over \pi}\ell n({x_{01}\over \rho})(Y-y)\}
dy Z(\b b_0+ {\b
x_{12}\over 2},\b x_{20},y,u) Z(\b b_0-{\b x_{20}\over 2},\b
x_{12},y,u),\eqno(12)$$

\noindent where the $\rho$ subscript on the integral means $x_{02} > \rho,
x_{12} >
\rho$ is
required.  $\rho$ is assumed small.  Eq.(12) can be formulated in an operator
language by
introducing $a$ and $a^+$ having commutation relation

$$[a(\b b,\b x), a^+(\b b^\prime,\b x^\prime)] = \delta(\b b-\b
b^\prime)\delta(\b x-\b x^\prime).\eqno(13)$$

\noindent  We introduce $ a^+(\b b,\b x)$ as an operator which creates a
dipole of separation $\b x$ and whose center  is at $\b b.$

The basic vertex in eq.(12) changes one dipole into two.  To that end we
introduce the
operator vertex $V_1$ by

$$V_1={\alpha N_c\over 2\pi^2}\int {x^2\over y^2z^2}\delta(\b x + \b y + \b z)
d^2bd^2xd^2yd^2z\  a^+(\b b+{\b y\over 2},\b z)
a^+(\b b-{\b z\over 2},\b y)a(\b b,\b x),\eqno(14)$$

\noindent a vertex which destroys a dipole and creates two dipoles.  The
integral in (14) is
restricted to $x_{02}, x_{12} > \rho,$ but we drop the $\rho$ subscript for
simplicity.
Also, introduce an interaction, $V_2$, giving virtual corrections by

$$V_2= - {2\alpha N_c\over \pi}\int d^2b d^2x \ell n(x/\rho) a^+(\b b,\b
x) a(\b b, \b x).\eqno(15)$$

\noindent Then

$$Z(\b b_0,\b x_{01},Y,u) = (0\vert e^{a_u}e^{YV(a^+,a)}
a^+(\b b_0,\b x_{01})\vert 0)\eqno(16)$$

\noindent where $V=V_1 + V_2$ and

$$a_u=\int d^2b d^2x u(\b b,\b x) a(\b b,\b x).\eqno(17)$$

\noindent The ``state'' $\vert 0)$ is such that $a(\b b,\b x)\vert 0) =0.$

In order to get a little more familiar with the operator language let us show
that
Z=1 when $u(\b b,\b x) \equiv 1$  Noting that

$$e^{a_u} a^+(\b b,\b x) e^{-a_u} = a^+(\b b, \b
x) + u(\b b,\b x),\eqno(18)$$

\noindent it is clear that

$$Z(\b b_0,\b x_{01},Y,u) = (0\vert e^{YV(a^++u,a)}
a^+(\b b_0,\b x_{01})\vert 0) + u(\b b_0,\b x_{01}).\eqno(19)$$

\noindent But

$$V(a^+ + u,{\bigg a})=\int d^2 bd^2 x\biggl\{{\alpha N_c\over 2\pi^2}
\int {x^2\over y^2z^2}\delta(\b x + \b y + \b z) d^2y d^2z$$
$$\biggl[a^+(\b b +{\b y\over 2},\b z)a^+(\b b-{\b
z\over 2},\b y) a(\b b,\b x)+a^+(\b b+{\b y\over 2},\b z)
a(\b b,\b  x) u(\b b-{\b z\over 2},\b y)$$
$$+ a^+(\b b-{\b z\over 2},\b y) a(\b b,\b x) u(\b b +{\b y\over
2}\b  z) + a(\b b,\b x) u(\b b + {\b y\over 2},\b  z) u(\b b-{\b z\over 2},\b
y)\biggr]$$
$$-{2\alpha N_c\over \pi}\ell n(x/\rho)[a^+(\b b,\b x) a(\b b,\b x)-u(\b b,\b
x) a (\b
b,\b x)]\}.\eqno(20)$$

\noindent When u=1 the term linear in a $a$  cancels meaning that $V(
a^++1,a)$ has $a^+
a^+ a$ and
$a^+ a$ terms only.  This gives  Z=1 in (19).

In order to make connection with the conventional approaches we now calculate
the
dipole
number density

$${\delta\over \delta u(\b b,\b x)} Z(\b b_0,\b x_{01},Y,u)\vert_{u=1}={1\over
2\pi
x_{01}^2} n(x_{01},x,Y,\vert\b b-\b b_0\vert).\eqno(21)$$

\noindent The factor of $[2\pi x_{01}^2]^{-1}$ in (21) is to keep the same
normalization
as
we have used in (4) and (5).  Using (19) and (20) it is straightforward to
obtain

$${\delta\over \delta u(\b b,\b x)} Z\vert_{u=1}=(0\vert a(\b b,\b x)
e^{Y a^+ K a}a^+(\b b_{01},\b
x_{01})\vert 0)\eqno(22 )$$

\noindent where

$$a^+ K\ a=\int d^2b d^2x d^2x^\prime d^2\zeta
a^+(\b b + \b \zeta,\b x^\prime)K(\b \zeta,\b x^\prime,\b x){\bigg a}(\b b,\b
x)\eqno(23)$$

\noindent with

$$K(\b \zeta,\b x^\prime,\b x)={\alpha N_c\over 2\pi^2}\ {x^2\over x^{\prime
2}(\b x-\b
x^\prime)^2}[\delta(\b\zeta-{\b x+\b x^\prime\over 2})+\delta(\b \zeta + {\b
x+\b
x^\prime\over 2})]$$
$$-{2\alpha N_c\over \pi}\ell n(x/\rho)\delta(\b \zeta)\delta(\b x-\b
x^\prime).\eqno(24)$$

\noindent Write

$$a(\b b,\b x)=\sum_{n=-\infty}^\infty \int{d\nu\over (2\pi)^2} 4({n\over 2} +
i\nu)a_{n\nu}(\b w){d^2w\over x^2}E^{n\nu}(\b b+\b x/2-\b w,\b b-\b x/2-\b
w),\eqno(25)$$

\noindent where the $E^{n\nu}$ are the eigenfuctions of the two Casimir
operators of the
conformal algebra and are given by Lipatov[3].  The completeness of the E's
means

$$\bigl[a_{n\nu}(\b w), a^+_{n^\prime\nu^\prime}(\b
w^\prime)\bigr]=\delta_{nn^\prime}\delta(\nu-\nu^\prime)\delta^2(\b w-\b
w^\prime).\eqno(26)$$

\noindent Also,

$$\int d^2bd^2xK(\b b^\prime-\b b,\b x^\prime,\b x){E^{n\nu}(\b b+\b x/2  -\b
w,\b b-\b
x/2-\b w)\over x^2}$$
$$={2\alpha N_c\over \pi}\chi(n,\nu)
{E^{n\nu}(\b b^\prime + \b x^\prime/2 -\b w,\b
b^\prime-\b x^\prime/2-\b w)\over (x^\prime)^2}\eqno(27)$$

\noindent allows one to write

$$a^+K a={2\alpha N_c\over \pi}\sum_{n=-\infty}^\infty \int
d\nu d^2 w \chi(n,\nu)a_{n\nu}^+(\b w)a_{n\nu}(\b w),\eqno(28)$$

\noindent where

$$\chi(n,\nu)=\psi(1)-{1\over 2}\psi({\vert n\vert +1\over 2} + i\nu)-{1\over
2}\psi
({\vert
n\vert +1\over 2}-i\nu).\eqno(29)$$

\noindent In arriving at (28) we have used eqs.23 and A.16 of ref.3.  Using
(25), (26) and
(28) in (22) one finds

$$n(x_{01},x,Y,b)=\sum_{n=-\infty}^\infty 16 \int{d\nu\over (2\pi)^3}\
{d^2w\over
x^2}
(\nu^2+{n^2\over 4})
e^{{2\alpha N_c\over \pi}\chi(n,\nu)Y}$$
$$\cdot E^{n\nu}(\b b+{\b x\over 2}-\b w,\b b-{\b x\over
2}-\b w)
E^{n\nu^*}({\b x_{01}\over 2}-\b w,-{\b x_{01}\over 2}-\b w).\eqno(30)$$

\noindent For large  Y  only the n=0 term need be kept.  Using

$$E^{0\nu}(\b \rho,\b \rho^\prime) = \bigl[{\vert \b \rho-\b
\rho^\prime\vert\over \rho
\rho^\prime}\bigr]^{1+2i\nu}\eqno(31)$$

\noindent it is easy to see that (8) results when $x/b, x_{01}/b << 1.$

In the operator formalism it is easy to write down an expression for the
scattering
amplitude of two high energy onium states.  In the center of mass of the
colliding onium
states we introduce $a^+(\b b,\b x)$ as a creation operator for
right-moving dipoles in the right-moving onium state and $d^+(\b b,\b x)$ as a
creaton operator for left-moving dipoles in the left-moving onium state.
Defining  F  as
in
(2) and the S-matrix by S=1+F one has

$$S(Y,b,x_{01},x_{01}^\prime)= (0\vert e^{ a_1 + d_1}e^{ - f} e^{{Y\over
2}(V_L+V_R)}
d^+(\b b+\b  b_0,\b x_{01}^\prime)a^+(\b b_0,\b x_{01})\vert
0)\eqno(32)$$

\noindent as the S-matrix for dipole-dipole scattering at impact parameter b
where $V_R
=
V_1 + V_2$ with $V_1$ and $V_2$ given by (14) and (15), and where one obtains
$V_L$ from
$V_R$ by replacing $a$ and $a^+$ by  $d$  and
$d^+.$  f  is determined by requiring that $F^{(1)},$ as given in (3), results
from expanding
$e^{-f}$ to first order in  f.  One finds, from (2) and (21),

$$f = 4\pi\alpha^2 x_{01}^2(x_{01}^\prime)^2\int d^2b_1\  {d^2x\over x^2}\
{d^2x^\prime\over (x^\prime)^2} {d\ell\over \ell^3}\bigl[1-J_0(\ell
x))\bigr]\bigl[1-J_0(\ell x^\prime)\bigr].$$
$$\cdot  a^+(\b b_1,\b x)a(\b b_1,\b x) d^+(\b
b_1,\b x) d(\b b_1,\b x).\eqno(33)$$

\noindent Using

$$\int_0^\infty {d\ell\over \ell^3}\bigl[1-J_0(\ell x)\bigr]\bigl[1-J_0(\ell
x^\prime)\bigr]={1\over 4} x_<^2(1+\ell n\  x_>/x_<),\eqno(34)$$

\noindent with $x_<$ the smaller of  $x$  and $x^\prime}$ and with $x_>$ the
larger of
$x$
and $x^\prime,$ one obtains

$$f=\pi\alpha^2x_{01}^2(x_{01}^\prime)^2\int d^2b_1 {d^2x d^2x^\prime\over
x^2(x^\prime)^2}
x_<^2(1+\ell n\  x_>/x_<)a^+(\b b_1,\b x)a(\b b_1,\b x)$$
$$\cdot d^+(\b
b_1,\b x^\prime)d(\b b_1,\b x^\prime).\eqno(35)$$

\noindent Recall again, the essential assumption in (33) is that  the important
contributions to (32) come from $x,x^\prime << \vert\b b_1-\b b_0\vert,\vert\b
b_1-\b b-
\b
b_0\vert$ so that the centers of the dipoles $\b b_1$ in (33), can be taken to
be the same
value.  This assumption is valid so long as

$$\ell n^2 b^2/R^2 << 14\alpha N_c\zeta(3)Y/\pi.\eqno(36)$$

If one expands $e^{-f}$,in (32), in powers of \ f\ a multiple scattering series
for\ F\
emerges.  However, the multiple scattering series is not useful here to impose
unitarity
because the parts of the onium wavefunction which dominate two or more
scatterings are
not
the parts of the wavefunction which dominate the full\ F,\ nor are they parts
of the
wavefunction which carry a significant amount of probability.  That is, rare
parts of the
onia wavefunctions dominate the multiple scattering series of two and more
pomeron
exchange,
and so the asymptotic parts of this series do not carry very useful information
about an
actual scattering.  In the Appendix two pomeron exchange is discussed in some
detail
to illustrate the problem with using a multiple scattering series in QCD.  In
the next
section, we exhibit a toy model which also shows the problems with using the
multiple
scattering series.

In terms of a large $N_c$ expansion (32) looks a little peculiar since\ f\ is
of order
$1/N_c^2$ when one counts orders of $N_c$ for $\alpha N_c$ fixed.  The
expansion we
have in
mind is the following.  At leading order in $N_c$ one simply takes the
scattering
amplitude
to be $F^{(1)}$ given by (3), or equivalently, given by taking $e^{-f}-1
\approx - f$ in
(32).  So long as $F^{(1)}$, which is of order $1/N_c^2$, is small this is the
answer in
the
leading logarithmic approximation to the large $N_c$ limit.  However, when  Y
becomes
large we may view $\alpha^2 e^{(\alpha_P-1)Y}$ and $\alpha N_c$ as two separate
parameters
which are not necessarily small while $\alpha,$ or $1/N_c,$ are regarded as
small
parameters.  Thus, we go beyond the large $N_c$ expansion but only when terms
which
are
formally higher order in $N_c$, like $\alpha^2 e^{(\alpha_P-1)Y},$ are enhanced
by a
large
factor due to large Y.

Even after recognizing that one must go beyond the large $N_c$ limit at high
energy, that
is
that terms down by $1/N_c^2$ may  effectively compete with leading order terms
when
Y  is
sufficiently large, the question remains as to  which nonleading terms in $N_c$
should be
kept.  In (32) the onia wavefunctions have been evaluated at leading order in
$N_c$ while
the scattering between individual dipoles in the colliding onia, the\ f\ given
in (35), is
taken at order $1/N_c^2$ which is the leading order at which dipole-dipole
scattering
occurs.  The ${1\over N_c^2}$ in\ f\ is compensated by the values that the
operators
$a^+a$ and $d^+d$ take in the right-moving and left-moving
onia states respectively.  $a^+ a$ is, roughly, the number of
dipoles in the right-moving onium, and at large  Y  the number of dipoles is a
large
number.
A leading
${1\over N_c^2}$ correction to the onium wavefunction may also be compensated
by the
large
number of dipoles (gluons) in the onium  wavefunction, however, such
corrections only
affect
a $small\ fraction$  of the dipoles in the onium wavefunction.  In an
onium-onium
scattering
one is sensitive to the average number of dipoles present in the onium
wavefunction, and
this
quantity has ${1/N_c^2}$ corrections which are not enhanced.

We may phrase this in a slightly different way.   f\ in (35) is no longer small
when the
number of dipoles, $a^+a,$ of the right-moving onium and the
number of dipoles, $d^+d,$ of the left-moving onium are of order $1/\alpha.$
When $a^+a$ is of order of  $1/\alpha$ then the probability of a
pair-wise dipole interaction within, say, the right-moving onium also becomes
of order 1.
However, the probability of any $given$ dipole in the right-moving onium to
have a
nonleading
order $N_c$ interaction is of size $\alpha$ and so the wavefunction is not
strongly
modified
until one reaches the ``saturation'' regime[9] where $a^+a$
becomes of size
$1/\alpha^2.$
\vskip 10pt
\centerline{\bf 4.  A Toy Model}
\vskip 6pt

Before trying to understand (32) it may be useful to consider a toy model where
the
scattering amplitude has the same structure as (32), but where the problem is
made
simpler
by eliminating the transverse dimensions.  While (32) is claimed to be a good
high energy
approximation to onium-onium scattering in four-dimensional QCD, the toy model
which
we
consider here is not likely a good approximation to any high energy scattering
problem in
field theory.

To introduce the model we define creation and annihilation operators obeying

$$[a,a^+] = 1$$

\noindent in analogy with (13).  We define a vertex $V_1$ by

$$V_1 = \alpha\  a^+a^+a\eqno(37)$$

\noindent in analogy with (14) and

$$V_2 = - \alpha\ a^+a\eqno(38)$$

\noindent in analogy with (15).  Define  Z(Y,u) by

$$Z(Y,u) = (0\vert e^{ a u} e^{Y(V_1+V_2)}a^+\vert 0)\eqno(39)$$

\noindent with  u  now a parmeter rather than a function.  We note that
Z(Y,1)=1.  We
view
Z(Y,u) as a generating function for the square of a wavefunction exactly as was
the case
in
Sect.3.  Eq.(39) is easily solved by noting that

$${dZ\over dY} = - Z + Z^2\eqno(40)$$

\noindent and

$$Z(Y=0,u) = u\eqno(41)$$

\noindent give

$$Z(Y,u) = {u\over u+(1-u)e^{\alpha Y}}.$$

\noindent For later discussion it is useful to observe that

$$Z(Y,e^{-\lambda}) = (0\vert e^a e^{-\lambda a^+a} e^{Y(V_1+V_2)}
a^+\vert 0).\eqno(42)

\noindent The probability of having exactly  n  quanta in the wavefunction,
$P_n(Y),$ is
given by

$$P_n={1\over n!}{d^nZ(Y,u)\over d u^n}\big\vert_{n=0}=e^{-\alpha
Y}(1-e^{-\alpha
Y})^{n-1}.\eqno(43)$$

\noindent Using

$$\=n = \Sigma_{n=1}^\infty nP_n = e^{\alpha Y}\eqno(44)$$

\noindent  we see that

$$\=n P_n = (1-{1\over\= n})^{n-1}\eqno(45)$$

\noindent so that for $\=n$ large

$$\=n P_n \approx e^{-n/\=n}\eqno(46)$$

\noindent obeys KNO scaling.

We introduce the S-matrix in analogy with (32) as

$$S(Y,f) = (0\vert e^{a+d} e^{-f\alpha^2a^+ad^+d} e^{\alpha{Y\over
2}(a^+a^+a-a^+a+d^+d^+d-d^+d)}
a^+d^+\vert 0)\eqno(47)$$

\noindent where $[d,d^+]=1$ and $[a,d] = [a^+,d]=0.$  Here
$f\alpha^2$ represents an interaction probability between individual quantum in
the two
colliding systems.  To first order in  f

$$S-1= - f \alpha^2 \=n^2(Y/2) = - f \alpha^2 e^{\alpha Y}\eqno(48)$$

\noindent in close analog to (8) and (10).  In fact, one can simplify (47)
considerably by
writing

$$S(Y,f) = \sum_{n,m=1}^\infty e^{-\alpha^2 f nm}P_n(Y/2) P_m(Y/2).\eqno(49)$$

\noindent Thus,

$$S= e^{-\alpha Y} \sum_{n=1}^\infty {(1-e^{-\alpha Y/2})^{n-1}\over
e^{\alpha^2fn}-1+e^{-\alpha Y/2}}.\eqno(50)$$

\noindent For $e^{\alpha Y/2}$ large

$$ S \approx e^{-\alpha Y} \sum_{n=1}^\infty {e^{-n/\=n (Y/2)}\over
e^{\alpha^2fn}-1+\=n(Y/2)^{-1}}.\eqno(51)$$

\noindent When $f\alpha^2 e^{\alpha Y}$ is small one can easily check that (51)
leads to
(48).  When $f\alpha^2 e^{\alpha Y}$ is large but $f\alpha^2$ is small one
finds

$$S \approx {1\over f\alpha^2} e^{-\alpha Y} {\rm min}\bigl\{\alpha Y - \ell
n(1/f\alpha^2),\ell n(1/f\alpha^2)\bigr\}.\eqno(52)$$

An   eikonal form for S, using   the right-hand side of (48) as the eikonal,
would lead to a
decrease of  S  as $exp\{-f\alpha^2 e^{\alpha Y}\}.$  Although  S  does indeed
go to zero
at
large  Y  that decrease does not follow an eikonal behavior.  It is instructive
to
evaluate the multiple scattering series for  S  from (49).  Formally,

$$S(Y,f) = \Sigma_{N=0}^\infty {(-f\alpha^2)^N\over N!}
\bigl[\Sigma_{n=1}^\infty n^N P_n (Y/2)\bigr]^2.\eqno(53)$$

\noindent For $\=n(Y/2)$ large one is tempted to use

$$\Sigma_{n=1}^\infty n^N P_n(Y/2) \approx {1\over \=n} \Sigma_{n=1}^\infty n^N
e^{-n/\=n}
\approx N![\=n(Y/2)]^N\eqno(54)$$

\noindent which would lead to

$$S(Y,f) = \Sigma_{N=0}^\infty [-f\alpha^2\=n^2(Y/2)]^N N!\eqno(55)$$

\noindent The series in (55) is a divergent series which, however, may be
defined by a
Borel
transform

$$S(Y,f) = \int_0^\infty db\  e^{-b/f\alpha^2}  \~S(Y,b).\eqno(56)$$

\noindent Eq.(55) would give

$$\~S(Y,b) = {1\over 1+b\=n^2(Y/2)}\eqno(57)$$

\noindent which is not quite right.  A more exact evaluation of the series
(53), not making
the high energy approximation (54), leads to

$$\~S(Y,b) = {1\over 1+\=n(e^{b\=n}-1)}\ \sum_{i=1}^\infty \delta({b\=n\over
f\alpha^2}-i).\eqno(58)$$

\noindent When $f\alpha^2 << [\=n(Y/2)]^{-1}$(58) reduces to (57), however when
$f\alpha^2
\geq [\=n(Y/2)]^{-1}$ one must use (58) in order to obtain the correct
asymptotic
behavior
given by (52).

There are several lesson which may be drawn from this calculation.  (i)  In our
toy model
the multiple scattering series does contain sufficient information to
reconstruct the
S-matrix using (56).  (ii)  To reconstruct the S-matrix  from the multiple
scattering
series requires a very precise evaluation of the individual terms in that
series.  The
leading high energy behavior for each term in the series is not sufficient as
the failure of
(57) to generate (52) shows.  (iii)  In our toy model the eikonal approximation
does not
give the correct high energy behavior of  S,  and it cannot be expected to work
any better
in QCD either.
\vskip 10pt

\centerline{\bf 5.  A Qualitative Picture of Unitarity at High Energies}
\vskip 6pt

In this section, we shall use (32) to estimate the energy at which the S-matrix
becomes
small for onium-onium scattering at an impact parameter  b.  When  S(Y,b) is
much less
than
one we shall speak of the scattering as being black.  As we shall see blackness
 sets in at
energies where the fixed coupling approximation is still valid, at least for
sufficiently
heavy onium states.  The approximations made in obtaining (32) remain valid
through the
energies where blackness sets in.  Thus the imposition of unitarity to obtain
blackness
occurs in the domain of perturbation QCD, a result which may at first sight
seem
surprising.  As one goes to even higher energies, at a fixed impact parameter,
saturation
effects bcome important and (32) ceases to be an accurate representation of the
onium-
onium
interaction.

Consider the S-matrix element given in (32).  The S-matrix for onium-onium
scattering is
obtained by integrating (32) over the heavy quark part of the onia
wavefunctions

$$S(Y,b)=\int d^2x_{01}d^2x_{01}^\prime \int dz_1dz_1^\prime
\Phi(x_{01},z_1)\Phi(x_{01}^\prime,z_1^\prime)S(Y,
b,x_{01},x_{01}^\prime).\eqno(59)$$

\noindent In the following, we shall deal with $S(Y,b, x_{01},x_{01}^\prime)$
with
$x_{01}$
and
$x_{01}^\prime$ taken to be 2R with\   R\ the average onium radius. (Of course
when\
Y\
becomes large  the average values of $x_{01}$ and $x_{01}^\prime$ in (59) may
be
much less
than\ R,\ but when that is the case the onium-onium scattering amplitude is
already quite
black.  Our purpose here is to find out at what energy blackness sets in, for a
given
impact parameter,b.)  We also suppose that $b/R >>1$ but that (36) is also
satisfied.  We
may formally write (32) as

$$S(Y,b) = \Sigma_{n,m} e^{-fnm}P_n(Y/2,\b b_0,\b x_{01})P_m(Y/2, \b b + \b
b_0,\b
x_{01}^\prime)\eqno(60)$$

\noindent where

$$P_n(Y/2,\b b_0,\b x_{01})=(n\vert e^{Y/2 V_R} a^+(\b b_0,\b
x_{01})\vert 0),\eqno(61)$$

\noindent with a similar expression for $P_m,$ and where

$$f_{nm} = (n m\vert f \vert n m).\eqno(62)$$

\noindent n labels an arbitrary state created from some number of creation
operator of the
type $a^+(\b b,\b x)$ with\ m\ being a state made of
$d^+'s.$  The operator\ f\ is given in (35) and is a diagonal operator in a
dipole number representation.  From (35) it is clear that $f_{nm}\geq 0.$  The
$P_m$ are
probability distributions obeying

$$\Sigma_n P_n (Y/2,\b b_0,\b x_{01}) = 1.\eqno(63)$$

\noindent We take $x_{01} = x_{01}^\prime = 2R$ and suppress the  R  dependence
in
S(Y,b) in
(60).  Since $f_{nm} >0$

$$S(Y,b) \geq exp\{- < f > \},\eqno(64)$$

\noindent where

$$<f> = \Sigma_{n,m} P_n P_m f_{nm} = - F^{(1)}(Y,b)\eqno(65)$$

\noindent with $F^{(1)}$ given by (9).  Thus, S(Y,b) cannot differ from 1 by
very much
until
$-F^{(1)}(Y,b)$ becomes sizeable.

When $\alpha N_cY$ is fixed and $\alpha N_c$ is small $S(Y,b) \approx 1-
F^{(1)}(Y,b)$ with
$F^{(1)}(Y,b)$ given by (9) and representing the leading logarithmic series.
As Y is
increased and $F^{(1)}(Y,b)$ grows the leading logarithmic series begins to
break down.
{}From (A.10) one  can see that $F^{(2)}(Y,b) = {1\over 2} < f^2>,$ the second
term in the
multiple scattering series becomes of order 1 before $F^{(1)}$ becomes of order
1.
However,
our understanding of the enhancement of $F^{(2)}$ as compared  to $[F^{(1)}]^2$
is
that that
enhancement is due to rare configurations in the wavefunction (see appendix)
and so does
not
have inportance for $<e^{-f}>.$  That is, a configuration which has a small
value of
$P_nP_m$ in (60), but which has a large value $f_{nm}$ contributes strongly to
$<
f^2>$ but
very weakly to $<e^{-f}>.$  We feel the value of $F^{(2)}$ is not very
interesting for the
unitization of the S-matrix because $F^{(2)}$ is sensitive to low probability
configurations.  On the other hand, $F^{(1)}$ appears not to be sensitive  to
low
probability
configurations and so we expect $<e^{-f}>$ to be significantly less than unity
when and
only
when $<f>$ is significantly  difficult from zero.

Let Y(b) be the value of Y, for a given  b, for which $-F^{(1)}(Y,b)=1$ with
$x_{01}=x_{01}^\prime = 2R.$  From (10) one finds

$$Y(b) = {1\over \alpha_P-1}\{\ell n{1\over 4\pi\alpha^2} + \ell n\  {b^2\over
R^2} - \ell
n
\ell n b^2/R^2 + {3\over 2} \ell n [{7\over 2}\alpha
N_c\zeta(3)Y(b)]\}.\eqno(66)$$

\noindent Eq.(66) does not quite give a solution for Y(b) since Y(b) also
appears on the
right-hand side of that equation.  However, the dependence of the right-hand
side of (66)
on
Y(b) is quite weak so that (66) is a useful way to represent Y(b).  Also, we
would not
expect
(66) to be accurate at the level of ${1\over \alpha N_c}$ terms in Y(b), and we
include ,
for example, the $4\pi$ in $\ell n({1\over 4\pi\alpha^2})$ only because that
factor
emerged
naturally.

In our toy model the expression analogous to (66) is $Y_0={1\over\alpha} \ell n
{1\over
f\alpha^2}$ which follows from (48).  If  we write $Y=Y_0 + \lambda/\alpha$
then from
(52)
one sees that $S(Y_0+\lambda/\alpha) = \lambda e^{-\lambda}$ so that  S  is
small when
$\alpha[Y-Y_0]>>1.$  We expect that to be the case for QCD also.  When $\alpha
N_c[Y-Y(b)]>>1$  we expect S(Y,b) to be small.  Proving , from (32) and (35),
that this
is
indeed the case would be very significant, however, so far we have been unable
to
construct a
solid argument to that end.

{}From (66) one sees that blackness sets in for b $\sim$ R at $Y(R) = Y_1,$
where

$$Y_1={1\over \alpha_P-1}\{\ell n({1\over 4\pi\alpha^2}) + {3\over 2} \ell
n[{7\over 2}
\alpha N_c\zeta(3)Y_1]\}.\eqno(67)$$

\noindent$Y_1$ is the smallest  Y  at which one expects unitarity corrections
to be
important.  At the other extreme when\   \ b  is as large as the diffusion
radius $\ell n^2
b^2/R^2 = {2\over a(Y)}$ with a given by (8) one obtains blackness at a
rapidity $Y_2$
with

$$Y_2 ={1\over \alpha_P-1}\biggl\{\ell n\bigl({1\over 4\pi\alpha^2}\bigr) +
{3\over 2}
\ell
n[{7\over 2}\alpha N_c\zeta(3)Y_2]+{\sqrt{14\alpha N_c\zeta(3)Y_2/\pi}}-\ell
n\ell
n(2/{\bigg a}(Y_2))\biggr\}.\eqno(68)$$

\noindent For $Y > Y_2$ (66) ceases to apply as the scattering is black over
the whole
region, in impact parameter. where partons have diffused.  A generalization of
(66) is
given
below for $Y > Y_2.$

Thus, we find it useful to view the scattering in three different energy
regimes.  (i)  When
$Y < Y_1$ the BFKL pomeron applies and $S(Y,b) = 1 + F^{(1)}(Y,b)$ with
anonium-
onium total
cross section

$$\sigma(Y) = - \int d^2b 2F^{(1)}(Y,b) =
16\pi^2R^2\alpha^2{e^{(\alpha_P-1)Y}\over
[{7\over
2}\alpha N_c\zeta(3)Y]^{1/2}}.\eqno(69)$$

\noindent (ii)  When $Y_1< Y < Y_2$  the S-matrix is black for

$$0 \le b^2 \le b^2(Y) = R^2 4\pi\alpha^2 {e^{(\alpha_P-1)Y}\ell n b^2/R^2\over
[{7\over
2}\alpha N_c\zeta(3)Y]^{3/2}} e^{-{a\over 2}\ell n^2\ ({b^2(Y)\over R^2})}$$

\noindent  and the BFKL pomeron applies for larger impact parameters.  One
finds a
cross
section

$$\sigma(Y) = \int_0^{b^2(Y)}2\pi d b^2 - \int_{b^2(Y)}^\infty 2\pi d b^2
F^{(1)}(Y,b)\eqno(70)$$

\noindent or

$$\sigma(Y) = 16\pi^2R^2\alpha^2 {e^{(\alpha_P-1)Y}\over [{7\over 2}\alpha
N_c\zeta(3)Y]^{1/2}} e^{-{a\over 2}\ell n^2\ ({b^2(Y)\over R^2})}[1+ a\ell
n{b^2(Y)\over R^2}]\eqno(71)$$

\noindent with a given by (8).  (iii) When $Y > Y_2$ the S-matrix is black out
to the
diffusion radius

$$b_{diff} = R^2 e^{{\sqrt{14\alpha N_c\zeta(3)Y/\pi}}}.\eqno(72)$$

\noindent Indeed, one can follow\ b\ somewhat beyond the diffusion radius.  So
long as
$\ell
n({b^2\over x_{01}x_{01}^\prime}) \leq \epsilon {2\over a}$ with $\epsilon <<
1$
eq.(9) is
a correct representation of the leading logarithmic approximation and means
that the
BFKL
pomeron can be used to infer cross sections up to $Y=Y_3$

\noindent where

$$Y_3=[\alpha_P-1-{28\alpha N_c\zeta(3)\over \pi}\epsilon]^{-1}\{\ell n({1\over
4\pi\alpha^2}) + {3\over 2} \ell n[{7\over 2}\alpha N_c\zeta(3)Y_3]
-\ell n\ell n({2\epsilon\over a(Y_3)})\}.\eqno(73)$$

\noindent Then for $Y_2 < Y < Y_3$ one obtains $\sigma(Y)$ using a formula
identical
to
(70).  One finds exactly the same formula as (71).  We note that for $Y < Y_2$
the BFKL
formula applies for the total cross section even though unitarity corrections
have become
inportant for large rgions of impact parameter space.  This is the case because
$\sigma(Y)$
is dominated by impact parameters obeying $\ell n b^2/R^2 \sim 1/a(Y).$  It is
only when
$Y
> Y_2$ that sizeable correctons to the BFKL formula occur.

It becomes crucial to consider wavefunction corrections beyond our leading
logarithmic
large $N_c$ limit only when the order of $1/\alpha$ dipoles of a given size
overlap.  This
occurs at lowest values of  Y  for dipoles of size  R  within a distance  R  of
the origin
of the heavy quark-antiquark pair.

Using (8) and requiring $n(R,R,Y/2, R) = 1/\alpha$  one finds

$$Y_{sat}={1\over \alpha_P-1}\{\ell n(1/\alpha^2) + 3 \ell n[{7\over 2} \alpha
N_c\zeta(3)Y_{sat}]\}\eqno(74)$$

\noindent a value parametrically larger than that given by (67), the value for
blackness of
scattering at b=R.  (In obtaining (74) we have set $\ell n {b^2/x_{01}x$ equal
to 1 in (8).
Equation (74) is not  accurate at terms of size $(\alpha N_c)^{-1}$ in
Y_{sat}.)$  For a
dipole of size x at impact parameter  b  the  Y  necessary for saturation is at
least as
great as the value given in (74) plus the rapidity necessary for dipoles of
size\  x\  and at
impact parameter\ b\ to be first produced.  This diffusion rapidity, $\Delta
Y,$ is
determined by $a(\Delta) \ell n^2{b^2\over R x} = 1$ giving

$$Y_{sat}(b,x) \geq {1\over \alpha_P-1}\{\ell n 1/\alpha^2 + 3\ell n[{7\over
2}\alpha
N_c\zeta(3)Y_{sat}]+{\pi\over 7\alpha N_c\zeta(3)}\ell n^2({b\over R
x})\}\eqno(75)$$

\noindent which is parametrically larger than Y(b) given by (66).  This means
that
unitarity
corrections in high energy scattering become important at rapidities where
saturation
effects in the wavefunctions of the colliding onia can safely be ignored. Thus
unitarization is a perturbative phenomenon.

Finally, we come to the point of the use of fixed coupling thoughout our whole
discussion.
The natural coupling for onium-onium scattering is $\alpha(R)$  with\ R\ the
onium
radius.
However, we have been led to consider distances as large as\ b\ determined by
$\ell n
{b^2\over R^2}\sim \epsilon {\bigg a}(Y).$  (See just  below   eq.(72).)  This
means that

$${\alpha(b)-\alpha(R)\over \alpha(b)} \leq \epsilon \alpha^2(R)
N_c^2Y.\eqno(76)$$

\noindent The largest value of\ Y\ we have considered is of size ${1\over
\alpha N_c} \ell
n({1\over \alpha N_c})$ so that

$${\alpha(b)-\alpha(R)\over \alpha(b)} \leq \epsilon \alpha N_c\ell n({1\over
\alpha
N_c})\eqno(77)$$

\noindent which means that for $\alpha N_c$ small the coupling can be
considered fixed
over
the regions of\   Y\   and\   b\   necessary for studying the unitarity problem
in high
energy onium-onium scattering.

\vskip 10pt
\centerline{\bf Appendix}
\vskip 6pt

In this appendix the two pomeron exchange amplitude is calculated, again for a
definite
(large) impact parameter between the two colliding onium particles.  We
continue to
follow
the procedure given in ref.5 where double pomeron exchange was calculated in
momentum
space.  We define $n_2(x_{01}, Y, b_i, x_i,b_j,x_j)$  as the dipole pair
density for
dipoles
of size $x_i$ and $x_j,$ at impact parameter $b_i$ and $b_j$ from the center of
the
heavy
quark-antiquark pair.  The heavy quark and anatiquark have a separation
$x_{01}$ and
all
the gluons which make up the dipoles\ i\ and\ j\ have momentum fractions
greater than or
equal to $e^{-Y}.$  The equation for $n_2,$ an equation analogous to eq.(27) of
ref.5, is

$$n_2(x_{01},Y,b_i,x_i,b_j,x_j)
={\alpha N_c\over \pi^2}\int_R{x_{01}^2d^2x_2\over x_{12}^2 x_{02}^2}\int_0^Y
dy
{e^{-{2\alpha N_c\over \pi}\ell n({x_{01}\over
\rho})(Y-y)}n(x_{02},y,\=b_i,x_i)$$
$$\cdot n(x_{12},y,\~b_j,x_j)
+{\alpha N_c\over  \pi^2}\int_R{x_{01}^2d^2x_2\over x_{02}^2 x_{12}^2}\int_0^Y
dy
e^{-{2\alpha N_c\over \pi}\ell n({x_{01}\over
\rho})(Y-y)}n_2(x_{12},y,\~b_i,x_i,\~b_j,x_j).\eqno(A.1)$$

\noindent We rcall that the\ R\ on the integrals in (12) means $x_{12} \geq
\rho,
x_{02}\geq \rho.$

\noindent Also $\b{\=b}=\b b-{\b x_0+\b x_2\over 2}$ and $\b{\~b} =\b b- {\b
x_1+\b
x_2\over
2}$  expresses the fact that on the left-hand side of (A.1) the $\b b's$ are
measured from
the
point $\b x_0+\b x_1\over 2}$ while on the right-hand side of (A.1)$\=b$ and
$\~b$ are
measured from ${\b x_0+\b x_2\over 2}$ and from ${\b x_1+\b x_2\over 2}$
respectively.

In order to solve (A.1), in the large  Y  limit, one must integrate over values
of $\b x_2$
such that $b_i$ and $b_j$ differ considerably from $\=b_1$ and $\~b_j$.  This
makes an
exact large  Y  solution difficult to find.  However, we can determine the form
of the
solution for $n_2,$ but not its normalization, by considering only values of
$\b x_2$ for
which $b_i$ and $b_j$ do not differ much from $\=b_i$ and $\~b_j.$  To that end
we
suppose the form of $n_2$ to be

$$n_2(x_{01},Y,b_i,x_i,b_j,x_j)
={e^{2(\alpha_P-1)Y}\ell n(x_{01}/x_i)\ell n (x_{01}/x_j)
x_{01}^{2-\gamma}\over
[{7\over
2}\alpha N_c\zeta(3)Y]^3 x_ix_j}$$
$$\cdot\ \ exp\{-{a(Y)\over 2}(\ell n^2(x_{01}/x_i)+\ell
n^2(x_{01}/x_j)\}h(b_i,b_j).\eqno(A.2)$$

\noindent In order to motivate the form given in (A.2) we first recall the
formula (8) for
$n(x_{01},Y,b,x).$  If we now restrict ourselves to the region where $\ell
n^2(b/x_{01})
<<
a^{-1}(Y)$ while $\ell n^2(b/x_i)$ and $\ell n^2(b/x_j)$ are of order $
a^{-1}(Y)$ we see that the logarithmic prefactors and the logarithmic terms in
the
exponential in (A.2) come directly from the inhomogeneous term in (A.1).  The
same is
true
for the factors $(x_ix_j)^{-1}$ and the Y-dependence.  The only surprising
factor in
(A.2) is
$x_{01}^{2-\gamma}$ rather than the, perhaps, natural $x_{01}^2$ factor which
is
explicit in
the first term on the right-hand side of (A.1).  We are now about to detrmine
the value of
$\gamma.$

If we substitute (A.2) into (A.1) we immediately see that (A.2) can be a
solution of (A.1)
only if $\gamma >  0.$  The dimensions of $n_2$ are given by the
$(x_ix_j)^{-1}$ factor
so
that $h(b_i,b_j)$ must be of the form $h(b_i,b_j) = b_i^{-2+\gamma}
h(b_i/b_j).$  Thus,
when
$x_{01}/b_i$ and $x_{01}/b_j$ are very small, with $b_i/b_j$ of order 1 we may
look
for a
consistent determination of $\gamma$ by dropping the inhomogeneous term in
(A.1).
Substituting (A.2) into (A.1) and dropping the inhomogneous term one finds

$$x_{01}^{2-\gamma} = {1\over 2\pi[\ell n x_{01}/\rho + 4\ell n 2]} \int_R
{x_{01}^2
d^2
x_2\over x_{02}^2 x_{12}^2} x_{12}^{2-\gamma}.\eqno(A.3)$$

\noindent Recall that

$${1\over 2\pi} \int_R {x_{01}^2 d^2x_2\over x_{02}^2 x_{12}^2} f(x_{12})\equiv
\int  dx_{12}
\~K(x_{01},x_{12}) f(x_{12})\eqno(A.4)$$

\noindent where

$$\~K(x_{01},x_{12})\mathop_{\longrightarrow\atop\rho\to 0}
K(x_{01},x_{12})+
\ell n(x_{01}/\rho)\delta(x_{01}-x_{12})\eqno(A.5)

\noindent with  K  the usual BFKL kernel.  Using (A.4) and (A.5) in (A.3) along
with the
eigenvalue equation

$$\int K(x_{01},x_{12}) x_{12}^{2-\gamma} dx_{12} = [\psi(1)-{1\over
2}\psi(1-\gamma/2)-{1\over 2}\psi(\gamma/2)] x_{01}^{2-\gamma}\eqno(A.6)$$

\noindent one finds (A.3) is satisfied provided

$$\psi(1) - {1\over 2} \psi(1-\gamma/2)-{1\over 2} \psi (\gamma/2) = 4\ell n
2.\eqno(A.7)$$

\noindent There are two solutions of (A.7) in the region $0 < \gamma < 2,$ the
region of
$\gamma$ for which the integral on the right-hand of (A.3) makes sense.  One
solution
has
$\gamma$ a little greater than 0 while the other solution has $\gamma$
a little below 2.  We presume that the
correct solution, the one dictated by correctly including the inhomogeneous
term in (A.1),
is the one with $\gamma$ a little above 0, however we have not been able to
prove this.

This double scattering contribution to  F  is given by a formula analogous to
eq.(45) of
ref.5,

$$F^{(2)}={1\over 2!}\int d^2b_id^2b_jn_2(x_{01},Y/2,b_i,x_i,b_j,x_j)
n_2(x_{01}^\prime,Y/2,\vert\b b_i-\b b\vert,x_i^\prime,\vert\b b_j-\b
b\vert,x_j^\prime)$$
$$\cdot 4\pi^2\alpha^2{dx_idx_i^\prime\  d\ell\over
x_ix_i^\prime\ell^3}[1-J_0(\ell
x_i)][1-J_0(\ell x_i^\prime)]4\pi^2\alpha^2{dx_jdx_j^\prime\ d\ell\over
x_jx_j^\prime\ell^3}[1-J_0(\ell^\prime\ x_j)][1-J_0(\ell^\prime\
x_j^\prime)].\eqno(A.8)$$

\noindent As usual the integrations over $x_i,x_j,x_i^\prime,x_j^\prime,\ell$
and
$\ell^\prime$ are easily done to give

$$F^{(2)}={32\pi^2\alpha^4e^{2(\alpha_P-1)Y}\over ({7\over 2}\alpha
N_c\zeta(3)Y)^3}\bigl({x_{01} x_{01}^\prime\over b^2}\bigr)^{2-\gamma}\int
d^2b_id^2b_j(b^2)^{2-\gamma}h(b_i,b_j)h(b_i-b,b_j-b).\eqno(A.9)$$

\noinent The final integral in (A.9) gives an (undetermined) constant so one
may write

$$F^{(2)}(Y,b)={c\over 2!}{\pi^2\alpha^4 e^{2(\alpha_P-1)Y}\over ({7\over
2}\alpha
N_c\zeta(3)Y)^3}\bigl({x_{01}x_{01}^\prime\over b^2}\bigr)^{2-\gamma}={c\over
2!}\bigl({b^2\over
x_{01}x_{01}^\prime}\bigr)^\gamma\bigl[F^{(1)}(Y,b)\bigr]^2\eqno(A.10)$$

\noindent with\ c\ the unknown constant.  In order to determine\ c\ it is
necessary to
include the inhomogeneous term (A.1) which, in effect, means that one must
solve (A.1)
also
in the region where $x_{01},x_{01}^\prime$ and\ b\ are comparable.  We remind
the
reader that
(A.10) can only be utilized in the region ${x_{01}\over b}, {x_{01}^\prime\over
b} <<
1$
while $\ell n^2 b/x_{01}, \ell n^2 b/x_{01}^\prime << 7 \alpha
N_c\zeta(3)Y/\pi.$

In the case of single pomeron exchange we found that the dominant contribution
to the
forward scattering amplitude came from impact parameters much larger than the
radius of
the
colliding onia.  For two pomeron exchange this is not the case.  $F^{(2)}(Y,b)$
decreases
faster than $1/b^2$ for $b/R >> 1$ so that the dominant contribution for two
pomeron
exchange
to onium-onium scattering comes when  b  is of the order of  R.  This is the
explanation
of
the fact that the Y-dependence of the two pomeron contribution to the forward
scattering
amplitude, $A^{(2)},$ is weaker than that of $(A^{(1)})^2$ by a factor of
$(\alpha
Y)^{-2}.$  At fixed impact parameters the Y-dependence of $A^{(2)}$ and
$(A^{(1)})^2$ is the
same, however, referring to (A.10) we note that for $b/R >> 1, F^{(2)}$ is much
larger
than
$(F^{(1)})^2.$  Thus, even in impact parameter space the multiple scattering
series  does
not resemble those we are used to in potential scattering and in scattering off
large nuclei.
\vill
\eject

\centerline{\bf Acknowledgment}
\vskip 6pt

Part of this work was done while I was a visitor at the LPTHE in Orsay.  I wish
to thank
Michel Fontannaz and the LPTHE for their hospitality and support.
\vskip 10pt
\centerline{\bf References}

\item{[1]}					Ya.Ya. Balitsky and L.N. Lipatov,
Sov.J.Nucl.Phys.28 (1978) 822.
\item{[2]}					E.A. Kuraev, L.N.Lipatov and V.S. Fadin,
Sov.Phys.JETP 45 (1977) 199.
\item{[3]}					L.N. Lipatov, Sov. Phys.JETP 63 (1986)
904.
\item{[4]}					A.H. Mueller, Nucl. Phys. B415 (1994) 373.
\item{[5]}					A.H. Mueller and B. Patel, Nucl. Phys.B.
(To be published).
\item{[6]}					N.N. Nikolaev and B.G. Zakharov, KFA-
IKP preprint (January 1994).
\item{[7]}					N.N. Nikolav, B.G. Zakharov and V.R.
Zoller, KFA-IKP preprint (January 1994).
\item{[8]}					Z. Koba,H.B. Nielsen and P. Olesen,Nucl.
Phys. B40 (1972) 317.
\item{[9]}					L.V. Gribov, E.M.Levin and M.G. Ryskin,
Phys. Rep.100C (1983)1.
\item{[10]}				M. Li and C-I Tan, Brown-HET 932, 943 (1994).

\noindent

\end
\bye